**The Impact of Industrial Zone:**

**Evidence from China's National High-tech Zone Policy**

Li Han

Department of Business, University of Chinese Academic of Social Science

April 15, 2023

## Abstract

Based on the statistical yearbook data and related patent data of 287 cities in China from 2000 to 2020, this study regards the policy of establishing the national high-tech zones as a quasi-natural experiment. Using this experiment, this study firstly estimated the treatment effect of the policy and checked the robustness of the estimation. Then the study examined the heterogeneity in different geographic demarcation of China and in different city level of China. After that, this study explored the possible influence mechanism of the policy. It shows that the possible mechanism of the policy is financial support, industrial agglomeration of secondary industry and the spillovers. In the end, this study examined the spillovers deeply and showed the distribution of spillover effect.

*Keywords*: Industry zones, staggered difference-in-difference, spatial Durbin model, spillovers

# I.introduction

At present, China's economic development has shifted from high-speed development to high-quality development, and at the same time, the situation of economic development is constantly changing, and the way of developing Chinese economic over the last forty year since 1978 has been greatly challenged. In the new way of developing Chinese economic, scientific and technological innovation is a key variable, the original innovation capacity has become more and more important for scientific and technological innovation, in order to improve the original innovation capacity, the state has formulated many relevant policies, these policies have effectively promoted industrial innovation, among which, the national high-tech zone policy is one of the important ones.

The National High-tech Industrial Development Zone is a major policy made by the State Council to develop China's high-tech industries, change the industrial structure, promote the transformation of traditional industries and enhance international competitiveness. The national high-tech zone in China was firstly established in 1988, since then, the number of national high-tech zones has increased year by year, showing two peaks of the establishment of high-tech zones, the first peak is 1988-1992, because of the change of ways of economic development, and the other peak is the past 10 years, with the structural transformation of Chinese economic. Nowadays, China has approved the establishment of 173 national high-tech zones, which including the Suzhou Industrial Park sharing the same industrial policies like the national high-tech zones, and the tendency of establishment of high-tech zones over the years is shown in Figure 1.Figure1 It shows that the output of national high-tech zones, the number of high-tech enterprises, and the number of national high-tech zones have been increasing sharply during the last 30 years. Looking back at the development of the national high-tech zone in the past 30 years, it can be divided into three stages, which are 1988-2000, 2000-2010 and 2010 to the present. (Wang,2018) The first stage of entrepreneurship is mainly through the aggregation of production factors and investment attraction, to expand the scale and volume of the industry. After the first stage of development, the industrial park has had an initial construction, but its value chain is still low-end, and the second stage is on this basis, through the "five transformations" to make the industrial cluster from production-driven to innovation-driven transformation (Zhang,2011). To achieve the upgrading of the value chain of the industry, as China has become the world's second largest economy, the construction of the national high-tech zone has also entered the third stage, and the nature and scope of innovation have undergone a major change. The scope of innovation has become a kind of comprehensive innovation.

The 30-year construction of national high-tech zones has not only successfully completed the main purpose of the policy, but also produced many other benefits. Nowadays, the national high-tech zones are becoming the main way to promote innovation and regional economic development. Since the implementation of the national high-tech zone policy means a lot of costs, whether the national high-tech zone policy can improve the regional innovation directly determines whether the policy should continue. Also, the establishment of national high-tech zones is a good quasi-natural experiment to examine the ways to estimate the treatment effect. Due to these reasons, the research of the high-tech zones has become the hot topics in academic research in Chinese economic.

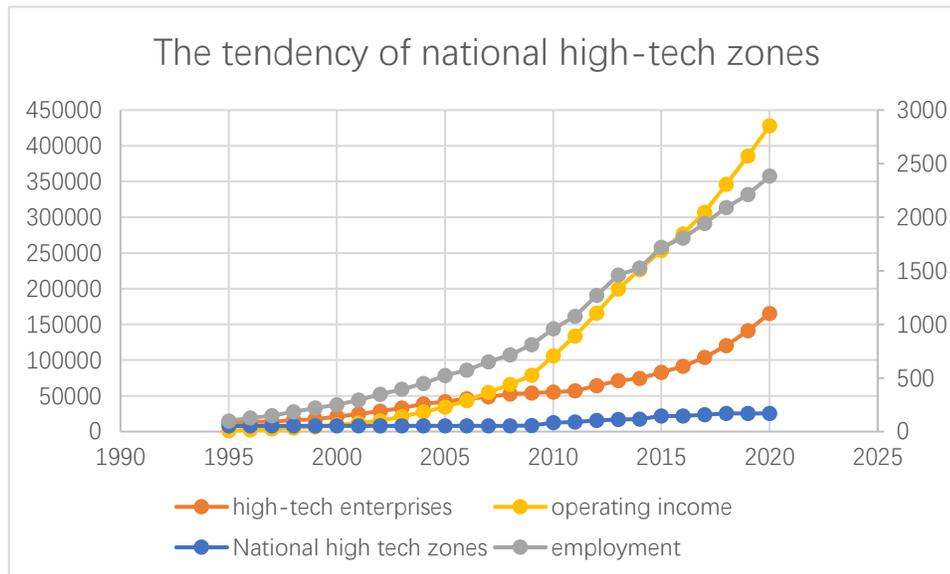

Figure1Establishment of national high-tech zones between 988 and 2022
Source: China Torch Statistical Yearbook

As for the impact of the establishment of high-tech zones, many researchers agreed that the establishment of national high-tech zones promotes economic growth and industrial innovation, and also tried to find how the policies worked. In terms of promoting economic growth, Tan et al. (2018) calculated that high-tech industrial parks can effectively achieve economic growth through technological progress and promote total factor productivity, and the effect of policies is greater in areas with a high degree of marketization. Han et al. (2021) analyzed the representatives of the high-tech zone in Zhongguancun, Beijing and explained the positive effect of the policy in regional innovation. Cheng et al. (2013) decompose the impact of total factor productivity through a stochastic frontier model to clarify the heterogeneity effect of policy on regional innovation. Díez-Vial (2017). examines the study of Spain's tech parks and found that even during recessions, companies inside the parks performed better than those outside the parks because they received fewer restrictions. In terms of the mechanism of the policy, many scholars have further analyzed the mechanism of action when analyzing the role of policy on regional innovation. Lin et al. (2018) through the analysis of enterprise data, believe that the role of development zones is to achieve higher production efficiency through a better policy environment. Xiong et al. (2019) pointed out that the mechanism of the high-tech zones is based on the original endowments. Xie et al. (2004) pointed out the specific role of industrial clusters in national high-tech zones from the perspective of industrial agglomeration and industrial clusters, Retinho (2010) showed the positive role of business incubators on innovation by examining the technology transfer and high-tech enterprises.

While the academic community recognizes the positive role of the national high-tech zones, some scholars also discuss the problems in the implementation of the national high-tech zone policy, Zhao et al. (2008) and Gu et al. (2015) respectively found through empirical evidence that high-tech industrial clusters do not achieve real clusters, but are simply piled up, and cannot give full play to their scope economic advantages. Cheng et al. (2003) also illustrate the corresponding problems in science and technology parks in their specific analysis of the Pearl River Delta region, Gao et al. (2019) divided the problems existing in the national high-tech zone in China into four parts: development purpose, regional layout, development level, and management system, so the heterogeneity of the national high-tech zone and the mechanism of policy need to be further studied.

In summary, the current research on the Chinese national high-tech zone has more focus on science and technology parks, but innovation data other studies have been used is as of 2015, the data in recent years is not widely used. And in recent years, the main purposes of establishing the national high-tech zones had changed a lot. Since 2015, the number of national high-tech zones that were established has reached 22, the number includes 12.7% of all current high-tech zones. The research method to examine the effect of the policy is the staggered difference-in-difference method. Although there are a lot of problems we should face to make the estimation exactly, Baker (2021) discuss the possible biases and solutions of the method. In view of this, this paper collects the innovation data of 244 prefecture-level cities from 2000 to 2020, uses the staggered difference-in-difference method to examine the effect of establishing the national high-tech zones on improving innovation ability, and further analyzes the heterogeneity of policy effects and the mechanism of policy action.

This study has various of possible marginal contribution. Firstly, the previous literature in Chinese national high-tech zone is mainly based on the Provincial-level data. City-level data is not widely used to clarify the impact of different city sizes on enterprise innovation. Secondly, the previous literature mostly uses staggered difference-in-difference model for research, and the advantage of staggered difference-in-difference model is that the time point of its processing is inconsistent so many policies that didn't happen in one time can be estimated. Also, the endogenous issue of policies, which usually happen if the time of treatment is same, caused by the traditional difference-in-difference method can be avoided. And due to China's national conditions, China often adopts the pilot rollout model in the process of policy formulation and implementation (Liu et al., 2015), and the staggered difference-in-difference method can effectively deal with this situation, but in recent years, scholars have expressed doubts about the effectiveness of this research method, some paper like Beck (2010) was argued by other researchers. And this paper uses Goodman-Bacon decomposition to enhance the robustness of the effect estimation of the national high-tech zone policy. Thirdly, the previous literature does not analyze the mechanism of action of national high-tech zone policy, and this paper studies the possible mechanism of action of national high-tech zone policy from three dimensions: financial support, industrial agglomeration and spatial spillover to analyze the mechanism of national high-tech zones to promote regional innovation.

## II.Literature review and analysis

### 1. The impact of national high-tech zones on regional innovation

The concept of innovation was first established in Schumpeter's (1912) book Economic Development Theory, which understood innovation as a way to establish a new production function and realize a new production combination to introduce new technologies, introduce new products, open up new markets, control the source of raw materials, and realize new organizational enterprises. On this basis, the concept of regional innovation systems gradually emerged. Lundvall (1985) proposed the concept of "innovation system", proposing that the flow of technology and information between labor, enterprises and governments is the key to the innovation process, and is the interaction between different participants. This concept was accepted and developed by other scholars. Freeman (1987) proposed the concept of a "national innovation system" in his study of the Japanese economy. The concept of innovation systems has become one of the important bases for explaining economic growth and formulating policies to promote economic development. On this basis, Cooke (1992) extended the concept of innovation system to the regional field by combining the phenomena of interregional linkages and knowledge spillover, and proposed the important concept of "regional innovation system". It is believed that the regional innovation system is mainly a regional industrial organization system composed of production enterprises and scientific research institutions related to geographical division of labor that can promote industrial innovation. The national high-tech zone in China is an industrial cluster that realizes industrial agglomeration and production factor agglomeration under the guidance of policies, and is a typical regional innovation system (Guo, 2011), and its evaluation methods mainly include productivity analysis method and production function method (Yuan, 2011). Previous studies have shown that the establishment of science and technology parks has a promoting effect on regional innovation capabilities. Löftten (2003) compared 273 enterprises located inside and outside science and technology parks in Sweden. Scott (2003) tested the development model of science and technology parks in the United States, and analyzed the relationship between American universities and science parks, and found that the establishment of science and technology parks can effectively promote the interaction between universities and enterprises. Fukugawa (2006) studied firms in science and technology parks in Japan and found that the establishment of science parks can effectively improve the value chain of enterprises. The establishment of national high-tech zones in China is usually accompanied by the implementation of a series of supporting policies, including policies such as talent introduction, land planning, capital subsidies, tax incentives, and financial support (Wang, 2011). The implementation of these policies has enabled the national high-tech zone to present a factor agglomeration and build an effective regional innovation system, and the agglomeration of these elements is finally transformed into effective innovation output in enterprises and universities and other scientific research institutions through the encouragement of local governments for innovation. In addition, although all national high-tech zones was established in the complete provincial-level cities and it only includes some part of the city, the agglomeration of industries and factors in national high-tech zones can produce large spatial spillover effects, so enterprises and universities in other areas of the city can also benefit from the establishment of national high-tech zone, and support the further development of the upstream industry of the national high-tech zone, so that the high-tech zone can play an innovation-leading role, and finally promoting regional innovation in the city. Because of this, the first problem that this thesis wants to research is:

H1: The policy of establishing national high-tech zones promotes regional innovation.

## 2. Heterogeneity of policy effects of national high-tech zones

Under the assumption that the effect of the national development zone policy on promoting economic growth and industrial innovation is evident, the size of its role must be limited by the gap between a lot of factors such as the initial endowment of the region. This gap results in obvious heterogeneity in the treatment effect of the national high-tech zone policy. Higher city level, higher government efficiency, worse factor market development, and more initial technical level is at both ends of the region all leads to more significant effect of the development zone policy (Zhou et al., 2018. Under different levels of development zones, the influence of national-level development zones is better than that of provincial-level development zones, but the impact of city-level and below development zones is not significant. Due to the obvious differences in the degree of economic development in China, regional economic development presents obvious imbalances, resulting in obvious differences in industrial structure between different regions, superimposed on other differences such as initial endowments and regional education levels (Xu et al., 2021; Huang et al., 2014; George et al., 2002), which makes China's regional development imbalance show obvious regional, urban-rural development imbalance and urban development inequality (An et al., 2018; Wang et al., 2018). Therefore, the implementation effect of the policy of the national high-tech zone will inevitably be different. Generally speaking, due to the high level of economic development and the initial endowment in the eastern part of China, so the establishment of the national high-tech zone in the eastern region can effectively promote the agglomeration of regional factors, while the central and western regions are limited by the level of development, the attraction of talents and capital is weak, and the financial support of local governments is low, and the ability to establish national high-tech zones to form industrial agglomeration and factor agglomeration in these regions is weak, and it is impossible to build a regional innovation system, so the impact of the national development zone policy on regional innovation is not obvious. At the city level, high-level cities are usually accompanied by larger urban scale (Wang et al., 2015), and larger urban scale means that it is easier to produce corresponding industrial agglomeration and promote the formation and development of regional innovation systems, so the national high-tech zones established in high-level cities have greater efficiency. To prove our analysis, there are two problems that we need to examine:

H2: The policy effect of the policy of establishing national high-tech zones is obviously heterogeneous at the regional level.

H3: The policy effect of the policy of establishing national high-tech zones is obviously heterogeneous at the city level.

## 3. Analysis of the mechanism of the national high-tech zone policy

The policy given in the National High-tech Zone usually includes a series of policies such as talent introduction, land planning, financial subsidies, tax incentives, and financial support and so on. (Lin et al.,2018) The mechanisms of action of these policies can be divided into two types of mechanisms: direct effect and indirect effect. On the one hand, the government's subsidies and support for related projects can directly change the choice tendency of enterprises and individuals in economic decision-making, so that more enterprises and individuals will invest more resources in scientific research and innovation, thereby improving the input of factors in the field of innovation and promoting regional innovation. Tian et al. (2018) proves that the flexible policy environment has effectively promoted investment and the construction and development of new enterprises, which in turn has promoted regional innovation. On the other hand, the government promotes the agglomeration of production factors, promotes the interaction and concentration of various factors, and promotes regional innovation through a series of policies such as talent introduction, and promotes industrial agglomeration and the formation of regional innovation systems. Brinkman et al. (2021) examined the high-tech zones in the United States and found that high-tech zones can bring regional innovation through industrial agglomeration, and the effects brought by different industrial agglomerations are also different. As Chang et al. (2021) and Wang et al. (2023) had shown, secondary industry agglomeration can promote regional economic development, while tertiary industry agglomeration plays a reverse role in economic development. Ye et al. (2023) showed that there is a significant correlation between manufacturing agglomeration and urban innovation performance U-shaped relationships. The manufacturing agglomeration of innovative cities has shown a significant positive effect. Also, financial support can effectively promote regional innovation efficiency (Wang et al.,2019). Financial support for high-tech enterprises in national high-tech zones is an important policy for state support for national high-tech zones.[1]Through the spatial spillover effect on the city, effectively promote the establishment and development of local high-tech enterprises, and then improve the efficiency of scientific and technological innovation. Xu et al. (2023) shows that science and technology finance policies can drive the improvement of China's urban green innovation level by increasing the government's fiscal expenditure on scientific and technological innovation and improving the efficiency of market credit fund allocation. Technology finance policies can also increase R&D investment by technology enterprises (Ren et al.,2023), Among the financial support, targeted lending support is one of the important support measures. At the same time, since innovative elements such as scientific research resources can flow between regions, the establishment of national high-tech zones can also promote the flow of resources from other regions to the region, Bernardí et al. (2007) use Spain 17 regions as examples. By using spatial econometric models, the existence of spatial spillover effects is proved. Moreno et al. (2004) use the data of 17 countries in the European Union and spatial models to estimate spillovers, and the results showed that the output of knowledge also appear to be influenced by other regions nearby. That influence is because of the spatial spillovers caused by patents and R&D in other regions. Such influence is not affected by national boundaries. Based on the three possible mechanisms that have been analyzed above, those three problems would be proved in the empirical study, that is:

H4: The policy of establishing national high-tech zones promotes regional innovation through industrial agglomeration

H5: The policy of establishing national high-tech zones promotes regional innovation through regional credit support.

H6: The policy of establishing national high-tech zones promotes regional innovation through spatial effects.

## III.Empirical studies

### 1. Baseline regression model setting

Since the establishment of national high-tech zones is carried out at different years, the establishment of national high-tech zones is a good quasi-natural experiment to evaluate the impact of China's industrial policy on regional innovation. As of 2022, a total of 173 state-level high-tech zones have been approved for establishment across the country, of which the latest one is the Karamay Park established in June 2022[2]. The specific difference-in-difference model setting is:

$$Y_{it} = \alpha_i + \lambda_t + \delta D_{it} + X_{it}\beta + \varepsilon_{it}$$

Among them, $t$ is the year, $i$ isfor the prefecture-level city, $Y_{it}$ is for the explanatory variable, that is the measure of regional innovation ability, $D_{it}$ is for the core explanatory variable of staggered difference-in-difference, $X_{it}$ is the control variable of the respective characteristics of the region, respectively, $\alpha_i, \lambda_t$ is the individual fixed effect and year fixed effect, $\varepsilon_{it}$ is for random error term, the size of $\delta$ is the coefficient of the core explanatory variable, that is, the size of the policy effect, if H1 is true then $\delta > 0$, and all variables with strong heteroscedasticity are logarithmic, and the missing values are generalized and imputed.

### 2. Variable selection

#### a) *Explained variable*

---

[2] These data was collected in the official site of Chinese government:
www.chinatorch.gov.cn/gxq/gxqmd/list.shtml

The explanatory variables in this document are the measurement of regional innovation. According to the innovation value chain theory, innovation can be divided into two stages: technology development and achievement transformation (Yu et al.,2009). In the technology development stage, its inputs are usually R&D manpower input and R&D capital investment, its output is usually patents, the input in the achievement transformation stage is the technical output and other factors obtained by the first stage of R&D, and its final output is the final income such as new product revenue. To simplify the research object and make sure that all the data can be collected properly, this research selects the number of patents application in the city-level admission boundary as the measurement of regional innovation. Due to the availability of the data, the time span of this research is from 2000 to 2019.

### b) *Core explanatory variable*

The core explanatory variable of this paper is the dummy variable $D_{it}$. Considering that the influence of the policy has a certain time interval, we assume that the policy of the Ministry of Science and Technology will take effect in the year after the establishment of the industrial park. From these assumptions, we construct the dummy variable $D_{it}$, if and only if the national high-tech zone has been established in the year of t and the region of I. Since the data in this paper is the panel data of prefecture-level cities, and some prefecture-level cities have established more than one national high-tech zone, it means that some individuals will be treated multiple times. Considering that the mechanism of high-tech zone policy includes policy overlay (Zhao et al.,2020) and spatial spillover effect on other areas of the city, this paper ignores the establishment of multiple treatment in a single city. We assuming that the treatment effect brought about by the second treatment is included in the first treatment, without considering the treatment effect brought about by the second treatment of the same prefecture-level city and the county-level city. With these assumptions, the coefficient is the treatment effect of the national development zone policy.

*c)  Moderator variables*

Combined with the existing literature, analysis H4 and H5 given above and the implementation characteristics of the national development zone policy itself, this paper draws on Liu et al. (2022) and Chang et al. (2021). We use the regional loan balance and the location entropy of the secondary and tertiary industries as the measurement of financial support and industrial agglomeration.

(1)  **Financial support (fin).** Existing studies usually use indicators such as the proportion of loans to local GDP (Huang et al.,2021), the concentration of financial resources (Liu et al., 2022), and the proportion of M2 to GDP (McKinnon et al.,1973). Since the difference-in-difference model requires that the treatment group must be randomly selected, on the basis of satisfying the exogenous nature of the treatment group selection, the size of financial support usually has nothing to do with the local economic development status, so combining the characteristics the policy itself and data sources, we use the loan balance of the prefecture-level city as a proxy variable for local financial support.

(2)  **Industrial agglomeration (lq).** The existing research measures of industrial agglomeration mainly include Gini coefficient, Herfindahl index, EG index, diversification index, etc. The location entropy is an important indicator to measure the degree of industrial agglomeration (Hansen et al.,2008). This paper draws on the methods of Yang et al. (2013) and Wang et al. (2023), considering the availability of data in various regions obtainability, using regional secondary industry and tertiary industry location entropy as a measure of industrial agglomeration, the specific calculation method is as follows:

$$lq_{ir} = \frac{e_{ir} \big/ \sum_i e_{ir}}{\sum_i e_{ir} \big/ \sum_r \sum_i e_{ir}}$$

Among them, the $e_{ir}$ number of employees in r industry in region $i$.

### d)  *Control variables*

**(1) Infrastructure (inf).** Infrastructure has a positive effect in promoting regional innovation, and the higher the infrastructure, the stronger the innovation ability of regions (Zhu et al. 2019). We use the measurement of infrastructure in Sheng et al. (2021) and Zhao et al. (2023) to used regional road passenger traffic to measure the level of infrastructure construction.

**(2)  Industrial structure (ind2, ind3).**Existing literature shows that industrial structure upgrading has a promoting effect on regional innovation (Zhao et al., 2018), and the national high-tech zone policy itself will play a significant role in promoting the amount of advanced industrial structure, that is, the three major industries gradually transition from the dominant position of the primary industry to the dominant position of the secondary industry and the tertiary industry (Zhu et al., 2018). For its measurement methods, there are secondary production and three production ratio, Thiel index and other methods (Xu et al., 2015), Considering that the changes in the ratio of secondary and tertiary industries reflect the upgrading of regional industrial structure, this paper controls the proportion of secondary and tertiary industries in GDP, and takes all the two variables as control variables.

**(3) Foreign Investment (fdi).** The use of foreign capital is one of the important factors in China's economic development. Especially since China's accession to the WTO in 2001, foreign investment has played an important role in regional innovation. The larger the amount of foreign investment, the stronger the role of promoting regional innovation, and the promotion role of foreign investment is quite different (Hou et al., 2006) But there is also a potential negative impact of foreign investment on the improvement of regional core technological innovation capabilities, core technologies are often difficult to obtain (Li et al., 2008). In this paper, referring to the practice of Kong et al. (2022), the actual amount of foreign funds is used as the control variable.

**(4) Research expenditure (rdcost) and scientific staff (rdperson).** Innovation input is an important factor affecting innovation output, and in terms of R&D input, R&D personnel and R&D expenditure are the two important innovation input indicators (Cao et al., 2012). Guan et al. (2010) researched, considering the comparability of data at the prefecture-level city level, so the expenditure on scientific research and the number of employees in the scientific research and technical services industry in this paper measure the local investment in scientific research material capital and human capital.

**(5) Level of regional economic development (lngdp).** There is a certain correlation between regional economic growth and regional innovation level, and it has stable coupling in the long term (Li et al., 2018), which is an important variable affecting regional innovation and has nothing to do with the establishment of national high-tech zones, and the important impact of national high-tech zones is that they could bring economic growth (Liu et al.,2015), so the level of regional economic development is an important control variable, so the logarithmic value of regional GDP is used as the control variable.

**(6) Local Government Fiscal Intervention (lngov).** In China, the government usually use some non-market action to do the macro adjustment and intervention of economics. Local government financial intervention has a great impact on regional innovation (Xu et al., 2015), and the measurement indicators for local government financial intervention include Fan Gang's marketization index, the proportion of local government fiscal expenditure to GDP and so on. In this study, we use the ratio of local fiscal expenditure to GDP to measure the effect of local government intervention.

**(7) Regional Informatization (lnict).** The digital economy significantly promotes high-quality development (Zhao et al., 2020), so the degree of informatization is an important influencing factor of innovation. This study draws on the research of Wan et al. (2022) and uses the number of regional mobile phones as the control variable.

In summary, combined with the definition of industrial policy by Lin (2018), this paper selects infrastructure, the proportion of added value of tertiary industry to GDP, foreign investment, scientific research expenditure, scientific research sector employment, local economic development level, local government financial cadre, and local informatization degree as the control variables.

## 3. Data

In China, the most authoritative data was issued in the National Bureau of Statistics of China. This institution usually issues a lot of data in the province level, and the data of each cities was not given officially. Because of it, we need to collect data in the statistic yearbook. We examined every statistic yearbook by using the CEInet Statistics Database, which shows all the data issued by the statistics bureau of each cities and provinces. Besides, the data of patents applications is gained in the Chinese Research Data Services Platform (CNRDS). If there is any missing value in the database, we will find the data from other data sources or make linear interpolation to maximize available data. But we do not use extrapolation because it will cause many unreasonable data like negative GDP. After collecting all the data, we matching the data by the name, the code of region and the year. Then we use the definition of all the control variables to generate the control variable. Because we want to examine the spillovers, we also collected geographic information in China. We use the geographic information from internet which is not official because official geographic information does not provide vector diagrams. Then we use the AutoNavi Open Platform to find the geographic coordinate of the national high-tech area. After getting the geographic location of these areas, we combine the location of the areas with the map of the boundary of the prefecture-level cities by using the ArcGIS software. After that, we generate the spatial weights matrix and the distance dummies by using the model we will show below.

The final descriptive statistics for each variable are Table11.

**Table1 Descriptive Statistics for Variables**

| variable name | N | mean | standard deviation | minimum | maximum |
|---|---|---|---|---|---|
| patentapplied | 6888 | 3821.17 | 12934.28 | 1.00 | 239892.00 |
| hightech | 6888 | 0.25 | 0.43 | 0.00 | 1.00 |
| lq2 | 5689 | 0.24 | 0.14 | 0.02 | 0.74 |
| lq3 | 5689 | 0.37 | 0.10 | 0.06 | 0.74 |
| lnfinance | 4855 | 25.14 | 1.34 | 21.76 | 29.72 |
| Inf | 5867 | 18.84 | 58.99 | 0.03 | 3482.55 |
| Ind3 | 6400 | 38.02 | 9.39 | 8.50 | 83.87 |
| Ind2 | 6397 | 46.30 | 11.38 | 10.68 | 90.97 |
| lnfdi | 5394 | 3.63 | 1.71 | 0.00 | 7.99 |
| rdcost | 6474 | 44205.29 | 252050.41 | 0.00 | 5550000.00 |
| rdperson | 5857 | 9270.45 | 34866.48 | 100.00 | 717100.00 |
| lngdp | 5891 | 10.06 | 0.98 | 7.40 | 13.19 |
| lngov | 5919 | 8.10 | 1.15 | 5.19 | 11.60 |
| lnict | 5373 | 0.49 | 0.32 | 0.00 | 2.41 |

# IV.Analysis of empirical results

In this part, we will first do the baseline regression by removing the high-tech zone established before 2000 due to the difference of the three stages of establishing the national high-tech zone. Then we will verify the availability of using difference-in-difference model by doing logit test, parallel trend. After doing these, we will use the Goodman-Bacon decomposition to further verify the correctness of removing them because these variables are the always treated group and it is the bad control group. In the end, we will do the placebo test and robustness check to prove the correctness of our findings. After having finished doing these baseline regressions, we will focus more on the heterogeneous and mechanism.

## 1. Baseline regression

Since there are significant differences between the three stages of establishing the national high-tech zone, this study excludes the data of national high-tech zones before 2,000 years of approval, which is in the first stage. After deleting the sample of high-tech zones that have been established before 2,000 years, we do the baseline regression using the difference-in-difference model and the results is shown in the table2. the core explanatory variables in regression (1) showed strong positive significance when using provincial clustering standard errors, which showed significant impact on regional innovation when the establishment of national high-tech zones was established. In regression (2)-(4), although control variables, individual fixed effects and year fixed effects were added, the strong positive significance does not change, which proves that the analysis

H1 is valid, that is, the establishment of national high-tech zones has a significant role in promoting regional innovation.

**Table3 baseline Regression**

| | (1)<br>patentapplied | (2)<br>patentapplied | (3)<br>patentapplied | (4)<br>patentapplied |
|---|---|---|---|---|
| hightech | 8648.0*** | 2829.8*** | 5874.9*** | 2500.5*** |
| | (2389.5) | (855.1) | (1838.3) | (861.4) |
| inf | | -27.86* | | -29.40* |
| | | (16.31) | | (15.98) |
| Ind2 | | 64.36 | | -1.774 |
| | | (43.29) | | (66.35) |
| Ind3 | | -73.00 | | -92.88 |
| | | (60.93) | | (81.54) |
| lnfdi | | -216.2 | | -252.5 |
| | | (181.3) | | (258.0) |
| rdcost | | 0.0363*** | | 0.0357** |
| | | (0.0131) | | (0.0137) |
| rdperson | | 0.0489 | | 0.0795 |
| | | (0.0624) | | (0.0672) |
| lngdp | | 3427.8* | | 2384.8 |
| | | (2014.2) | | (2205.5) |
| lngov | | -2303.6* | | -2157.6 |
| | | (1390.6) | | (1280.8) |
| lnict | | 6156.2* | | 5111.8 |
| | | (3509.0) | | (3177.4) |
| city fixed effect | No | No | Yes | Yes |
| time fixed effect | No | No | Yes | Yes |
| _cons | 1663.3*** | -14429.7* | 188.8 | -3305.9 |
| | (478.9) | (7558.6) | (785.3) | (13847.2) |
| r2 | | | 0.207 | 0.571 |
| r2_w | 0.150 | 0.567 | 0.207 | 0.571 |
| r2_b | 0.109 | 0.653 | 0.109 | 0.605 |
| r2_o | 0.123 | 0.615 | 0.139 | 0.587 |
| r2_a | | | 0.204 | 0.568 |

t statistics: $^*p < 0.1$, $^{**}p < 0.05$, $p < 0.01$, and the standard error uses robust standard error.

## 2. Suitability test for double difference

Since the formulation and implementation of China's policies are usually not selected randomly, and the selection of the policy often according to a certain standard to select pilot areas and then promote the method of gradual implementation. The the difference-in-difference method requires that the treatment group must be randomly selected, so we must make sure that all the zones was selected randomly before regression analysis. Because policy adoption and non-adoption is a binary variable, this paper refers to the method of Guo(2018).The logit model was used to test the randomness of the establishment of national development zones on the original data and the deletion of data from prefecture-level cities that had established national high-tech zones 2,000 years ago, and the results are listed in Table4:

| **Table4Self-selection test for samples** | | | |
|---|---|---|---|
| (1) | (2) | (3) | (4) |
| hightech | hightech | hightech | hightech |
| Deletion | No | No | Yes | Yes |
| patentapplied | 0.00015*** | -0.00001 | 0.00011*** | 0.00002 |
| | (0.00002) | (0.00001) | (0.00001) | (0.00001) |
| inf | | -0.00173 | | -0.00660 |
| | | (0.00125) | | (0.00457) |
| Ind2 | | 0.0946*** | | 0.0543*** |
| | | (0.0098) | | (0.0124) |
| Ind3 | | 0.0314*** | | 0.0179* |
| | | (0.0079) | | (0.0104) |
| lnfdi | | -0.0418 | | -0.0788 |
| | | (0.0385) | | (0.0499) |
| rdcost | | -0.00000* | | 0.00000* |
| | | (0.00000) | | (0.00000) |
| rdperson | | 0.00004*** | | -0.00003*** |
| | | (0.00001) | | (0.00001) |
| lngdp | | 2.475*** | | 1.926*** |
| | | (0.194) | | (0.243) |
| lngov | | -0.850*** | | 0.662*** |
| | | (0.130) | | (0.151) |
| lnict | | -1.009*** | | -1.926*** |

|        |         | (0.318)    |          | (0.547)   |
|--------|---------|------------|----------|-----------|
| _cons  | -1.570***  | -24.27***  | -2.227*** | -29.10*** |
|        | (0.0447)  | (1.227)    | (0.0533)  | (1.833)   |
| r2_p   | 0.140   | 0.363      | 0.118    | 0.337     |

t statistics: $^{*} p < 0.1$, $^{**} p < 0.05$, $p < 0.01$, the first two groups did not delete bad controls, the latter two groups had removed samples that are always processed groups, and the standard error uses robust standard error.

As shown in the table above, regressions (1) and (2) are cases where data from prefecture-level cities in the first stage that have been treated before 2,000 years ago are not deleted, and regressions (3) and (4) shows the regression that all the prefecture-level cities which were treated before 2000 are deleted. As can be seen from the regression (2) and (4) above, the insignificance of the patent applied shows that the self-selection has been significantly decreased after the treatment group of the first stage of entrepreneurship is deleted. At the same time, in order to make the above deletion more credible and rationality and make the result more robustness, we test whether the above two groups data meet the parallel trend hypothesis and the difference in regression coefficients, this is shown in Figure2:

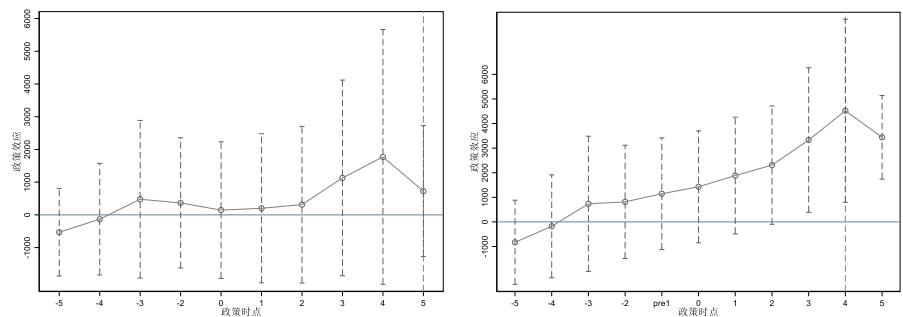

**Figure2 Parallel trend test**

As can be seen from the results of the parallel trend test plotted in figure 2, the parallel trend test does not pass. When the sample of development zones established before 2,000 years is not included, the parallel trend test passed.

Without removing the development zones established 2,000 years ago, the baseline regression results are shown in the table5below. The results of the regression without removing the development zone that were established before 2000 are shown, and it can be seen from (1) and (2) that when the removal is not happened, the significance of the treatment effect is totally ignored. And it can be seen from the regression (1)(3)(4) shows that the estimated coefficients of the core explanatory variables are also not significant when the region and time fixed effect is added to the model.

**Table6Regression results without some samples**

|  | (1) | (2) | (3) | (4) |
|---|---|---|---|---|
| Removal | No | No | Yes | Yes |
| hightech | 2242.1 | 1249.9 | 5874.9*** | 2500.5*** |
|  | (2322.5) | (1237.9) | (1838.3) | (861.4) |
| inf |  | -17.27*** |  | -29.40* |
|  |  | (5.031) |  | (15.98) |
| Ind2 |  | -57.85 |  | -1.774 |
|  |  | (91.93) |  | (66.35) |
| Ind3 |  | -138.2 |  | -92.88 |
|  |  | (95.80) |  | (81.54) |
| lnfdi |  | -160.3 |  | -252.5 |
|  |  | (243.8) |  | (258.0) |
| rdcost |  | 0.0350*** |  | 0.0357** |
|  |  | (0.00399) |  | (0.0137) |
| rdperson |  | 0.139*** |  | 0.0795 |
|  |  | (0.0329) |  | (0.0672) |
| lngdp |  | 3148.0 |  | 2384.8 |
|  |  | (2463.1) |  | (2205.5) |
| lngov |  | -3215.5* |  | -2157.6 |
|  |  | (1618.6) |  | (1280.8) |
| lnict |  | 4974.2** |  | 5111.8 |
|  |  | (2141.7) |  | (3177.4) |
| _cons | -150.3 | -1272.6 | 188.8 | -3305.9 |
|  | (958.5) | (15463.0) | (785.3) | (13847.2) |
| r2 | 0.173 | 0.766 | 0.207 | 0.571 |
| r2_w | 0.173 | 0.766 | 0.207 | 0.571 |
| r2_b | 0.182 | 0.735 | 0.109 | 0.605 |
| r2_o | 0.130 | 0.739 | 0.139 | 0.587 |
| r2_a | 0.170 | 0.764 | 0.204 | 0.568 |

t-Statistics: $^*$ $p < 0.1$, $^{**}$ $p < 0.05$, $p < 0.01$, standard error uses the clustering standard error at the provincial level.

Using the method of Andrew Goodman-Bacon (2021), the treatment effect of regression (4) in the above table is divided into five parts, and the decomposition results [3]are shown in Table7. According to the decomposition results, it can be seen that the estimation of the control group as "always treated" is significantly negative, and the cities that have established development zones themselves already contain the policy effect of development zones, and due to the heterogeneity of processing around 2000, they are the bad control groups, in order to reduce the estimation bias in this part, it is necessary to delete the "always treated" part, which means deleting the sample of cities that established development zones before 2000. From the decomposition, we show that deleting data before 2000 is reasonable.

**Table7 Goodman-Bacon decomposition**

| category | Estimates | weight |
|---|---|---|
| Timing groups | 318.9360394 | 0.179087985 |
| Timing vs always treated | -14195.31157 | 0.203117354 |
| Never treated vs timing | 6678.120077 | 0.601346382 |
| Never treated vs always treated | -330238.3125 | 0.000102218 |
| Within group | 13308.70898 | 0.016346062 |

## 3. Robustness test

---

[3] This paper is written here using a the stata package "bacondecomp". The decomposition can be carried out and the control group and the treatment group can be output at the same time, which is relatively clearer.

The problem that the use of the double difference method for policy evaluation may be that a certain policy may be superimposed with other policies, or caused by other random factors, and the estimated treatment effect may not be caused by the national high-tech zone policy, but by other policies carried out at the same time. To rule out the possibility, we use the placebo test (Zhou et al.2018) method, the implementation year of the policy and the implementation object is randomized, the pseudo-experimental group was constructed for the same regression analysis. The progress is repeated 500 times, and the distribution of coefficient of the core explanatory variables were plotted in Figure3:

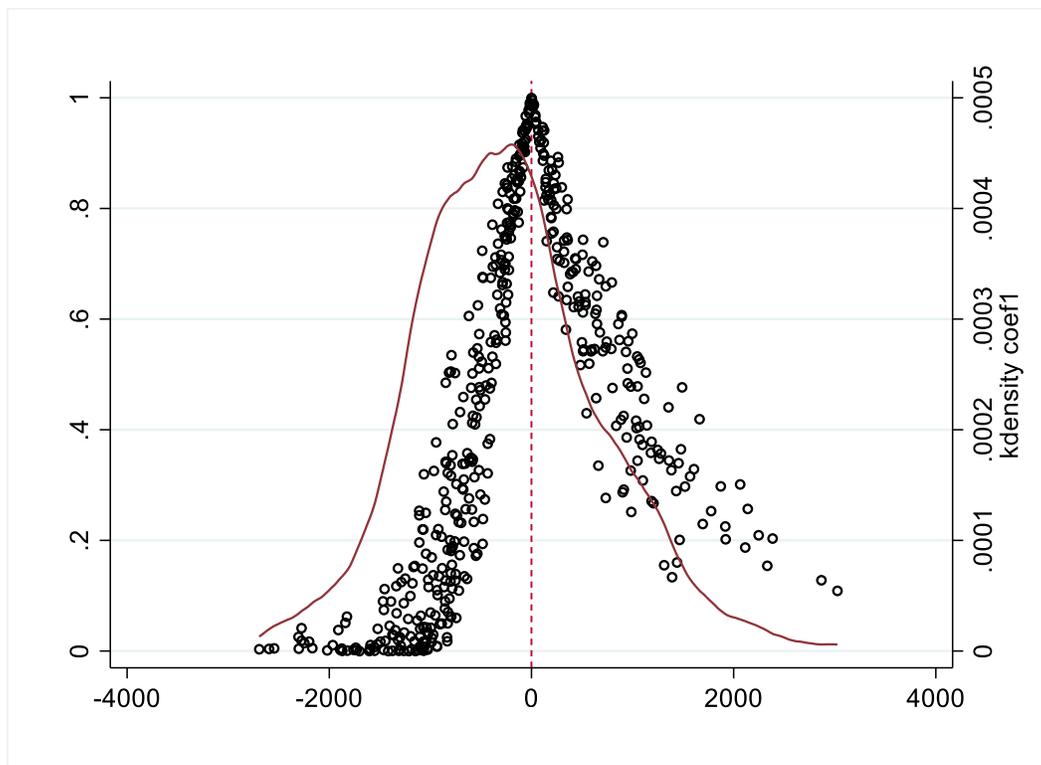

**Figure3 Placebo test**

According to the estimation results after repeated simulation of 500 times in the figure above, it can be seen that the regression coefficient fluctuates around 0, which is significantly not equal to the estimated value of the benchmark regression of 2500.5, which means the placebo test has passed. Passing the placebo test means that the effect of the policy is not due to other policies or random factors over the same period.

To make the result more robustness, we also replace the control variables. Due to the official reply of the cycle of applying for a new patent, there is usually a lag of 1.5-3 years from patent application to acquisition[4], Due to these reason, we use the number of second-order lagged variable of patent acquisition to replace the number of patent applications, and make the baseline regression again, and the conclusions were broadly similar, as shown Table8, demonstrating that this result was relatively robust.

**Table8 Robustness tests**

|  | (1) patentgained | (2) patentgained | (3) patentgained | (4) patentgained |
|---|---|---|---|---|
| hightech | 7559.7*** | 2231.7*** | 4995.0*** | 2059.1** |
|  | (2059.2) | (828.7) | (1554.3) | (848.7) |
| inf |  | -21.70 |  | -24.24 |
|  |  | (15.80) |  | (16.24) |
| Ind2 |  | 98.26** |  | 29.57 |
|  |  | (38.47) |  | (55.52) |
| Ind3 |  | -40.39 |  | -78.27 |
|  |  | (43.85) |  | (67.88) |
| lnfdi |  | -176.1 |  | -237.2 |
|  |  | (143.0) |  | (194.0) |
| rdcost |  | 0.0273*** |  | 0.0271*** |
|  |  | (0.00816) |  | (0.00827) |
| rdperson |  | 0.0361 |  | 0.0447 |
|  |  | (0.0355) |  | (0.0305) |
| lngdp |  | 2634.6* |  | 1624.6 |
|  |  | (1444.8) |  | (1654.9) |
| lngov |  | -1988.2* |  | -1605.5 |
|  |  | (1115.5) |  | (1039.4) |
| lnict |  | 6756.1* |  | 5451.2* |
|  |  | (3628.2) |  | (3190.9) |

---

[4] See https://new.qq.com/rain/a/20201205A01GKN00

| | (1) | (2) | (3) | (4) |
|---|---|---|---|---|
| _cons | 1623.6*** | -12105.1** | 196.8 | -1584.0 |
| | (495.0) | (6006.4) | (679.0) | (12866.0) |
| r2 | | | 0.209 | 0.506 |
| r2_w | 0.146 | 0.495 | 0.209 | 0.506 |
| r2_b | 0.104 | 0.633 | 0.104 | 0.565 |
| r2_o | 0.116 | 0.578 | 0.132 | 0.538 |
| r2_a | | | 0.205 | 0.502 |

t Statistics: * $p < 0.1$, ** $p < 0.05$, $p < 0.01$, use the city-level clustering standard error.

## 4. Heterogeneity test of policy effects of national high-tech zones

In order to further explore the differences in the treatment effects of development zone policies in different regions, this paper further examines the heterogeneity of this treatment effect at the regional and city levels to test the analysis H2 and H3.

Firstly, through the regional level heterogeneity test H2, this paper uses the division standard of the National Bureau of Statistics in China, and divides the data into four major regions: eastern, central, western and northeastern and the results was shown in table9. regression (1)-(4) from Table10illustrated that the coefficient of the core explanatory variable in the eastern region is positive and significant, and in the central region, the coefficient is positive, small and significant under the significance of 10%, in the western region, the coefficient is negative and not significant. Because the northeast region has only three provinces, the to exclude that the effect is not significant due to the small sample size in the northeast region. In summary, the policy implementation effect is strongest in the eastern region, the second in the central region, the weakest in the western region, which indicates that the analysis H2 is valid.

**Table10 Tests for regional heterogeneity**

| (1) | (2) | (3) | (4) |
|---|---|---|---|
| Eastern | Central | Western | Northeastern |

| | | | | |
|---|---|---|---|---|
| hightech | 3358.9** | 619.1* | -269.8 | 31.39 |
| | (1498.9) | (324.3) | (494.7) | (230.8) |
| inf | -60.20*** | 2.233 | 1.588 | 3.765 |
| | (12.33) | (3.226) | (1.809) | (4.594) |
| Ind2 | -443.2*** | 22.28 | 31.23* | -23.03* |
| | (160.2) | (28.15) | (17.13) | (11.93) |
| Ind3 | -621.0*** | 61.23** | 1.943 | 3.059 |
| | (226.7) | (23.37) | (16.00) | (8.934) |
| lnfdi | -629.4 | 111.8 | -1.668 | -46.77 |
| | (802.9) | (107.5) | (42.78) | (27.72) |
| rdcost | 0.0296** | 0.0484*** | 0.135*** | 0.0798*** |
| | (0.0120) | (0.00184) | (0.0180) | (0.0221) |
| rdperson | 0.0772 | -0.0271 | 0.0985** | 0.00714 |
| | (0.0868) | (0.0258) | (0.0493) | (0.00604) |
| lngdp | 13003.1*** | -1051.5** | 176.0 | -367.8 |
| | (4808.4) | (485.2) | (427.9) | (352.9) |
| lngov | -7419.5* | 480.0 | -79.58 | -472.2 |
| | (3992.3) | (627.0) | (306.8) | (337.2) |
| lnict | 11503.9* | 848.0 | -460.1 | 406.9 |
| | (6130.6) | (761.2) | (418.6) | (716.3) |
| _cons | -26539.4 | 2786.4 | -2350.2 | 7130.8*** |
| | (24340.0) | (5821.6) | (3117.3) | (2550.0) |
| r2 | 0.669 | 0.819 | 0.812 | 0.735 |
| r2_w | 0.669 | 0.819 | 0.812 | 0.735 |
| r2_b | 0.649 | 0.644 | 0.969 | 0.856 |
| r2_o | 0.652 | 0.774 | 0.905 | 0.763 |
| r2_a | 0.661 | 0.815 | 0.807 | 0.720 |

$t$ statistics: $^*$ $p < 0.1$, $^{**}$ $p < 0.05$, $p < 0.01$, because there are too few provincial areas, we use the city-level clustering standard error. [5]

Next, using the city level data, further check whether the analysis H3 is true, about the division of city level, there are many current classification standards in academia, the traditional division method is usually divided according to the size of the urban population, but the indicator of population size has certain limitations (Qi et al., 2016), the economic development conditions of the region and the level of regional economic development is totally ignored. Referring to the practice of Liu et al. (2019) and Han et al. (2021), and uses the annual list of top 100 cities launched by the New First-Tier Cities Research Institute[6] to divide the regions into third-tier cities above and below third-tier cities. The results are shown in Table11. The regression (1)-(4) show that the treatment effect of the high-level city group is positive and significant, and the sensitivity to policy is strong. On the contrary, the treatment effect of the low-level urban group is negative and weak, the sensitivity to policy is weak, and there is even a certain reverse effect, which proves that the analysis H3 is valid.

---

[5] SeeAngrist, J. D., & Pischke, J. S. (2008). Mostly harmless econometrics: An empiricist's companion. Princeton university press. Here the author gives a fuzzy criterion. When the number of clusters is approximately less than 42, the clustering criterion is no longer stable.

**Table11 City-level robustness tests**

|  | (1) high patentapplied | (2) low patentapplied | (3) high patentgained | (4) low patentgained |
|---|---|---|---|---|
| hightech | 791.5$^{***}$ | -6111.4 | 688.7$^{**}$ | -591.1 |
|  | (289.1) | (4378.3) | (268.6) | (3090.4) |
| inf | -0.757 | -28.66$^{*}$ | 1.050 | -47.99$^{***}$ |
|  | (1.954) | (14.30) | (1.852) | (11.90) |
| Ind2 | 28.78 | -3773.3$^{**}$ | 43.72$^{**}$ | -3667.4$^{**}$ |
|  | (18.09) | (1788.2) | (17.96) | (1529.9) |
| Ind3 | 10.80 | -3690.0$^{*}$ | 9.306 | -3509.1$^{**}$ |
|  | (17.07) | (1986.7) | (15.70) | (1654.5) |
| lnfdi | -34.24 | -2357.7 | -64.26 | -256.3 |
|  | (54.39) | (2678.9) | (49.94) | (1977.0) |
| rdcost | 0.0651$^{***}$ | 0.0203$^{***}$ | 0.0386$^{**}$ | 0.0149$^{***}$ |
|  | (0.0135) | (0.00689) | (0.0168) | (0.00325) |
| rdperson | 0.0417 | 0.0915 | 0.0216 | 0.0517 |
|  | (0.0263) | (0.0698) | (0.0170) | (0.0363) |
| lngdp | -212.2 | 42074.7$^{**}$ | -229.5 | 29855.0$^{**}$ |
|  | (397.6) | (17732.7) | (342.3) | (13050.6) |
| lngov | -38.65 | -10837.3 | -98.94 | -5072.2 |
|  | (320.5) | (7109.6) | (331.2) | (7381.1) |

---

[6] The official site is www.datayicai.com.

| | | | |
|---|---|---|---|
| lnict | 60.13 | -13186.4 | 44.17 | -4323.7 |
| | (774.3) | (14719.0) | (514.2) | (11910.8) |
| _cons | 422.4 | 13840.8 | 712.3 | 70283.8 |
| | (3580.9) | (138420.9) | (3432.2) | (114959.7) |
| r2 | 0.619 | 0.809 | 0.517 | 0.778 |
| r2_w | 0.619 | 0.809 | 0.517 | 0.778 |
| r2_b | 0.490 | 0.417 | 0.269 | 0.285 |
| r2_o | 0.560 | 0.643 | 0.400 | 0.557 |
| r2_a | 0.616 | 0.789 | 0.513 | 0.755 |

*t* Statistics: $^{*}$ $p < 0.1$, $^{**}$ $p < 0.05$, $p < 0.01$, use the city-level clustering standard error.

## 5. Influence mechanism of the national high-tech zone policy

In order to further explore the influence mechanism of national high-tech zone policy, we use the moderation effect model on the basis of baseline regression, add the interaction terms of core explanatory variables and regulatory variables, and explores the possible mechanism of enterprise innovation in national high-tech zones to determine whether the analysis H4 and H5 are correct.

Firstly, the correctness of the analysis H4 is tested, and the number of Table12.

Table12 The moderating effect of financial support regression

| | (1) | (2) | (3) |
|---|---|---|---|
| | patentapplied | patentapplied | patentapplied |
| hightech | 2500.5$^{***}$ | 2211.2$^{***}$ | -150067.3$^{***}$ |
| | (861.4) | (770.8) | (31327.8) |
| lnfinance | | 294.9 | 10.64 |
| | | (1102.2) | (1006.2) |
| Interaction | | | 5883.9$^{***}$ |
| | | | (1211.3) |
| inf | -29.40$^{*}$ | -31.86$^{**}$ | -24.60$^{*}$ |
| | (15.98) | (14.70) | (12.26) |
| Ind2 | -1.774 | -65.79 | -44.77 |
| | (66.35) | (82.82) | (74.58) |
| Ind3 | -92.88 | -134.8 | -95.22 |
| | (81.54) | (109.2) | (91.44) |
| lnfdi | -252.5 | -241.6 | -193.0 |
| | (258.0) | (262.4) | (231.0) |
| rdcost | 0.0357** | 0.0329** | 0.0251** |
| | (0.0137) | (0.0123) | (0.0104) |
| rdperson | 0.0795 | 0.152** | 0.131* |
| | (0.0672) | (0.0624) | (0.0672) |
| lngdp | 2384.8 | 3362.4 | 3060.1 |
| | (2205.5) | (2571.0) | (2064.9) |
| lngov | -2157.6 | -2315.5 | -1817.8 |
| | (1280.8) | (1670.3) | (1372.4) |
| lnict | 5111.8 | 1536.6 | 597.7 |
| | (3177.4) | (1874.2) | (1706.9) |
| _cons | -3305.9 | -12698.5 | -9122.3 |
| | (13847.2) | (33991.7) | (29079.3) |
| r2 | 0.571 | 0.582 | 0.636 |
| r2_w | 0.571 | 0.582 | 0.636 |
| r2_b | 0.605 | 0.559 | 0.633 |
| r2_o | 0.587 | 0.566 | 0.633 |
| r2_a | 0.568 | 0.579 | 0.633 |

t-Statistics: $^*$ $p < 0.1$, $^{**}$ $p < 0.05$, $p < 0.01$, standard error uses the clustering standard error at the provincial level.

Regression (1) in the table is a baseline regression, regression (2) is a regression after adding a moderator, and regression (3) is a regression with an interaction term, from regression (1)-(3). It can be seen that when the regulating variable is added, the coefficient size of the core explanatory variable decreases but is still significantly positive, and when the modulator and the interaction term are added at the same time, the sign of the core explanatory variable has changed from positive to negative and the absolute value increases significantly. These regressions indicate that the financial support really have the effect on promoting the innovation. Combining the above regression results, it can be shown that the analysis H4 is valid.

Then, we analysis the influence of industrial aggregation. As we have analyzed above, the variable of location entropy of the secondary industry and the tertiary industry is used as moderator. Adding the moderator and interaction term, the results obtained are shown in Table Table13and t-Statistics: $^{*}$ $p < 0.1$, $^{**}$ $p < 0.05$, $p < 0.01$, standard error uses the clustering standard error at the provincial level.

**Table14**:

| | (1) | (2) | (3) |
|---|---|---|---|
| | patentapplied | patentapplied | patentapplied |
| hightech | 2500.5*** | 1452.7** | -9925.3*** |
| | (861.4) | (598.0) | (1658.5) |
| lq2 | | 27938.4*** | 17385.1*** |
| | | (8742.4) | (6259.9) |
| hightech_lq2 | | | 29218.6*** |
| | | | (4884.6) |
| inf | -29.40* | -21.23* | -14.57 |
| | (15.98) | (11.89) | (8.996) |
| Ind2 | -1.774 | -57.77 | -39.92 |
| | (66.35) | (51.14) | (46.76) |
| Ind3 | -92.88 | -126.1 | -103.2 |
| | (81.54) | (78.07) | (68.79) |
| lnfdi | -252.5 | -301.5 | -287.8 |
| | (258.0) | (266.6) | (240.9) |
| rdcost | 0.0357** | 0.0351** | 0.0349*** |
| | (0.0137) | (0.0127) | (0.0120) |
| rdperson | 0.0795 | 0.0960 | 0.104 |
| | (0.0672) | (0.0649) | (0.0632) |
| lngdp | 2384.8 | 1332.7 | 1124.4 |
| | (2205.5) | (1745.9) | (1558.4) |
| lngov | -2157.6 | -1045.1 | -1039.7 |
| | (1280.8) | (1173.6) | (1170.7) |
| lnict | 5111.8 | 5970.7* | 5891.0* |
| | (3177.4) | (2938.0) | (3033.7) |
| _cons | -3305.9 | -1309.7 | -86.52 |
| | (13847.2) | (9559.5) | (8768.2) |
| r2 | 0.571 | 0.606 | 0.623 |
| r2_w | 0.571 | 0.606 | 0.623 |
| r2_b | 0.605 | 0.665 | 0.687 |
| r2_o | 0.587 | 0.637 | 0.656 |

Table13of secondary industry agglomeration regressed

| | (1) | (2) | (3) |
|---|---|---|---|
| r2_a | 0.568 | 0.603 | 0.621 |

t-Statistics: $^*p < 0.1$, $^{**}p < 0.05$, $p < 0.01$, standard error uses the clustering standard error at the provincial level.

**Table14 the regulatory effect of tertiary industry agglomeration**

| | (1) | (2) | (3) |
|---|---|---|---|
| | patentapplied | patentapplied | patentapplied |
| hightech | 2500.5$^{***}$ | 2098.0$^{***}$ | 11508.0$^{***}$ |
| | (861.4) | (721.8) | (2954.1) |
| lq3 | | -24318.5$^{***}$ | -18242.2$^{***}$ |
| | | (7113.8) | (4851.7) |
| hightech_lq3 | | | -24475.9$^{***}$ |
| | | | (6790.4) |
| inf | -29.40$^*$ | -22.77$^*$ | -19.24$^*$ |
| | (15.98) | (12.78) | (10.95) |
| Ind2 | -1.774 | -70.62 | -62.73 |
| | (66.35) | (62.70) | (61.09) |
| Ind3 | -92.88 | -154.0$^*$ | -134.4 |
| | (81.54) | (89.41) | (80.00) |
| lnfdi | -252.5 | -277.2 | -259.0 |
| | (258.0) | (258.9) | (240.9) |
| rdcost | 0.0357$^{**}$ | 0.0353$^{**}$ | 0.0354$^{**}$ |
| | (0.0137) | (0.0132) | (0.0129) |
| rdperson | 0.0795 | 0.0894 | 0.102 |
| | (0.0672) | (0.0673) | (0.0665) |
| lngdp | 2384.8 | 2196.3 | 2082.1 |
| | (2205.5) | (1969.5) | (1822.4) |
| lngov | -2157.6 | -1687.2 | -1704.3 |
| | (1280.8) | (1166.7) | (1177.7) |
| lnict | 5111.8 | 5774.2$^*$ | 5540.8$^*$ |
| | (3177.4) | (2910.1) | (2857.3) |
| _cons | -3305.9 | 9390.9 | 6866.6 |
| | (13847.2) | (10238.2) | (9490.1) |
| r2 | 0.571 | 0.594 | 0.605 |
| r2_w | 0.571 | 0.594 | 0.605 |
| r2_b | 0.605 | 0.635 | 0.664 |
| r2_o | 0.587 | 0.616 | 0.635 |
| r2_a | 0.568 | 0.591 | 0.602 |

t-Statistics: $^*p < 0.1$, $^{**}p < 0.05$, $p < 0.01$, standard error uses the clustering standard error at the provincial level.

In Table 10, the regression (1) is the baseline regression, the regression (2) is the regression of the entropy of the secondary industry, and the regression (3) is the regression of the interaction item. Regression (1) and regression (2) showed that the coefficient size of the core explanatory variable decreased significantly after adding the control variable of location entropy, and the coefficient size of the core explanatory variable was significantly reduced after adding the interaction term, the coefficient of the interaction term is significantly positive, indicating that the secondary industry agglomeration is an important reason for the policy role of the national high-tech zone. In the same way, after adding the interaction term of the tertiary industry location entropy in Table 11, the coefficient of the core explanatory variable increases significantly, and the coefficient of the interaction term is significantly negative. That means that the agglomeration of the tertiary industry is not an important reason for the role of the national high-tech zone policy. It shows that the H5 is partly true. The agglomeration of the secondary industry is conducive to regional innovation, and the excessive agglomeration of the tertiary industry is not conducive to the effect of the national high-tech zone policy.

Next, we will use the spatial Durbin model and the ring methods to examine the spillovers, the spatial effect of the establishment of national high-tech zones on neighboring prefecture-level cities is discussed, and the specific model is as follows:

$$Y_{it} = \alpha_i + \lambda_t + \delta D_{it} + X_{it}\beta + \rho w_i y_t + w_i X_t \gamma + \varepsilon_{it}$$

Among them, $t$ is for the year , $i$ is for the prefecture-level city, $Y_{it}$ is for the explanatory variable, that is, the innovation ability, $D_{it}$ is for the core explanatory variable of staggered difference-in-difference, $X_{it}$ is the control variable of the respective characteristics of the region, respectively, $\alpha_i, \lambda_t$ is the individual fixed effect and the time fixed effect, $\varepsilon_{it}$ for the random error term, the size of $\delta$ is the coefficient of the core explanatory variable, that is, the size of the policy effect, $\rho y_t + w_i X_t \gamma$ is the spatial effect, where the first term is the spatial autocorrelation effect, the second term is the spatial spillover effect, $w_i$ is the $i$th row of the spatial weight matrix, $y_t$ represents the value of the explanatory variable in other regions, the size of $\rho$ represent the effect of spatial autocorrelation. The second term $X_t$ represents the value of the control variable in other regions, $\gamma$ is the influence of the control variable in other neighboring regions. If H6 is true then $\delta > 0$. To make the Spatial Durbin model, all missing values are interpolated. If the value that being interpolated is negative, we will replace them with zero to make sure that all the values are meaningful. Then, we use the adjacency matrix as $w_i$ to estimate the model, and the result is shown in Table15**Table16**:

**Table16 Panel Spatial Durbin Model**

|  | (1) | (2) | (3) |
|---|---|---|---|
|  | patentapplied | patentapplied | patentapplied |
| hightech | 3225.6*** | 2226.2*** | 2058.9*** |
|  | (721.2) | (620.7) | (626.7) |
| inf | -4.509 | -4.066 | -3.775 |
|  | (3.303) | (3.422) | (3.039) |
| Ind2 | -11.23 | -38.74 | -72.07* |
|  | (37.57) | (31.15) | (42.31) |
| Ind3 | -97.90** | -98.92*** | -129.3*** |
|  | (43.69) | (37.29) | (44.94) |
| lnfdi | -234.0 | -220.9 | -218.3 |
|  | (166.2) | (155.8) | (154.4) |
| rdcost | 0.0357*** | 0.0283*** | 0.0280*** |
|  | (0.00867) | (0.00788) | (0.00706) |
| rdperson | -0.0222 | 0.00442 | 0.00506 |
|  | (0.0301) | (0.0241) | (0.0291) |
| lngdp | 1856.1$^*$ | 1367.1 | 1931.6$^*$ |
| | (1085.5) | (944.5) | (1088.4) |
| lngov | -1415.1$^*$ | -1450.1$^{**}$ | -1463.3$^{**}$ |
| | (770.6) | (677.0) | (632.3) |
| lnict | 5058.5$^{**}$ | 4689.6$^{***}$ | 4396.7$^{**}$ |
| | (2122.6) | (1808.1) | (1852.8) |
| _cons | -2350.4 | | |
| | (6654.4) | | |
| $\rho$ | | 0.391$^{***}$ | 0.301$^{***}$ |
| | | (0.0586) | (0.0689) |
| r2 | 0.537 | 0.646 | 0.594 |
| r2_w | 0.537 | 0.590 | 0.613 |
| r2_b | 0.637 | 0.705 | 0.578 |
| r2_o | 0.586 | | |
| r2_a | 0.535 | | |

t-Statistics: $^*\,p < 0.1$, $^{**}\,p < 0.05$, $p < 0.01$, standard error uses the clustering standard error at the provincial level.

It can be seen from the regression results (2) and (3) that compared with the baseline regression (1), after adding spatial effects, the core explanatory variables were significantly positive, but their values were significantly reduced. The ρ value was positive and significant, which proved that there was a certain positive correlation between the existence of national high-tech zones and the innovation of the cities in the adjacent areas. This results initially verify the existence of spatial spillover effect and indicated that spatial spillover effect was a possible mechanism of national high-tech zone policy.

Next, using the ring method mentioned above, referring to the research methods of Lu (2015) et al., to test the spatial spillover effect to prove the correctness of the analysis H6. Also, we will use these methods to provide me with a distribution of the spillover effect. The model setting are as follows:

$$Y_{it} = \alpha_i + \lambda_t + \delta D_{it} + X_{it}\beta + \sum_{n=1}^{m} \gamma_n + \varepsilon_{it}$$

Compared with the basic model, the distance dummy variables $\gamma_n$ are added to the model to show the spillover effect, and the value of $\gamma_n$ is:

$$\gamma_n = \begin{cases} 1, \text{if there are more than one high} - \text{tech zones in the ring of } \big((n-1)d, nd\big) \\ 0, \text{others} \end{cases}$$

In order to ensure the robustness of the test, 10 km and 20km were used as $d$ values for the model. After adding the distance dummies to the baseline regression, the regression results are shown in Table17 and the distribution is shown in Figure 4.

Table18 Ring regression

| | (1)20km patentapplied | (2)10km patentapplied | (3)20km patentgained | (4)10km patentgained |
|---|---|---|---|---|
| hightech | 2418.0*** | 2257.4*** | 2118.7*** | 2056.9*** |
| | (637.4) | (643.6) | (643.1) | (654.1) |
| inf | -28.51* | -29.75** | -23.08 | -23.79 |
| | (14.84) | (14.95) | (15.96) | (16.13) |
| Ind2 | -52.57 | -57.29 | -6.335 | -13.92 |
| | (42.03) | (44.57) | (35.14) | (37.39) |
| Ind3 | -114.9** | -113.3** | -90.92** | -91.89** |
| | (50.63) | (47.69) | (40.12) | (38.98) |
| lnfdi | -246.7 | -253.9 | -221.2* | -221.5* |
| | (177.3) | (171.2) | (131.6) | (126.1) |
| rdcost | 0.0346*** | 0.0335*** | 0.0265*** | 0.0255*** |
| | (0.0120) | (0.0118) | (0.00724) | (0.00722) |
| rdperson | 0.0851 | 0.0926 | 0.0460* | 0.0520* |
| | (0.0584) | (0.0591) | (0.0275) | (0.0284) |
| lngdp | 1787.5 | 1609.0 | 1304.8 | 1203.2 |
| | (1189.1) | (1076.0) | (1001.0) | (947.9) |
| lngov | -2032.4* | -1927.0* | -1579.1** | -1434.0* |
| | (1074.2) | (1039.1) | (786.0) | (756.3) |
| lnict | 5848.2** | 6005.6** | 5851.5** | 5898.9** |
| | (2527.8) | (2573.4) | (2444.9) | (2464.3) |
| _cons | 1936.2 | 2896.4 | 1212.6 | 1527.7 |

| | (7136.7) | (7195.0) | (6575.8) | (6536.7) |
|---|---|---|---|---|
| *N* | 3964 | 3964 | 3755 | 3755 |

t-statistics: $^*$ $p < 0.1$, $^{**}$ $p < 0.05$, $p < 0.01$, standard error using the cluster standard error, 10 km and 20km are the *d* values mentioned above.

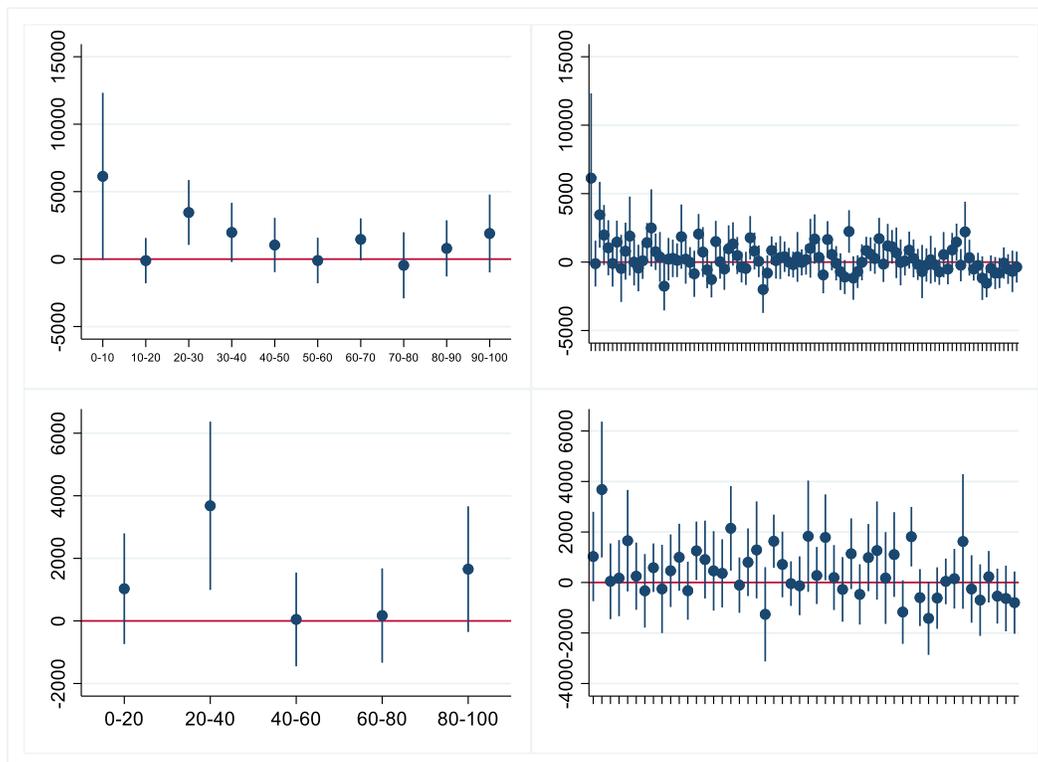

**Figure4 Spatial spillover effects**

From the above regression results and the graph, it can be seen that regardless of whether the ring radius is 10 km or 20km, when considering the spatial spillover effect of the adjacent high-tech zone, the core explanatory variables are still positive, and the dummy variables at different geographical distances are positive at the closer distance, and are numerically close to 0 at the longer distances[7], which further proves the validity of the analysis H6. That shows that there is an obvious spatial spillover effect in the policy role of the national high-tech zone, and the spatial spillover effect within the adjacent range is more obvious.

---

[7] In fact, the spillover happens only between 10km and 40km. When the distance grows over 40km, the spillover effect is not obvious because the coefficient of distant dummies fluctuate around zero.

## V.Conclusions and policy recommendations

In summary, this research uses the panel data of 287 cities in 2000-2019 to comprehensively estimate the impact of the establishment of national high-tech zones on regional innovation by regarding the establishment of national high-tech zones as a quasi-natural experiment. At the same time, the Bacon decomposition method is used to further prove the reliability of baseline regression and the sample selection. After that, the research investigates heterogeneity of the treatment effect on the city level and the region of China. The results show that the effect of the establishment of national high-tech zones has significant regional differences, except for the small sample in the northeast region, the policy implementation effect is the strongest in the eastern region, followed by the central region, the weakest in the western region, and even has a weak countereffect in the western region. At the city level, high-level cities have a strong sensitivity to policies, while low-level cities have a weak sensitivity to policies, and even have a certain reverse effect. Finally, this paper uses the moderating effect model and spatial Durbin model to explore the mechanism of national high-tech zone policy on regional innovation promotion, and the results show that industrial agglomeration, regional credit support, spillovers, etc. are the possible mechanisms of national high-tech zone policy.

This study provides some new evidence of the positive effect of the national high-tech zone. We combine the difference-in-difference methods with decomposition of Goodman-Bacon, spatial Durbin model and ring methods to prove the robustness of regression and the spillover effects. The conclusions of this paper have certain implications for policy.

First, in the stage when China's economic growth has shifted to high-quality growth, the national high-tech zone policy has a significant role in promoting industrial innovation and improving the original innovation ability, and the establishment of national high-tech zones in some areas can effectively promote local economic development and industrial innovation while this kind of policy doesn't work in many other regions in China. Second, the imbalance of development between regions still exists, limited by the initial endowment of the region, natural conditions and other factors, the sensitivity of each region to the policy is different. In the eastern region, the national high-tech zone policy can effectively promote regional development, while in the central and western regions, the effect of this policy is limited, and more effective targeted policies need to be formulated to promote local innovative development. Third, due to the imbalance of urban and rural development in China, the effect of this policy in small and medium-sized cities is restricted. It is necessary to take more resources to small and medium-sized cities, build urban agglomerations to promote the improvement of local initial development conditions, so as to enhance local innovation capabilities. Fourth, we should further understand the role mechanism of policies, take market-oriented as the guide, formulate more effective and precise policies such as targeted loans to play the role of policies more effectively.